\documentstyle[12pt]{article}

\def\overlay#1#2{\ifmmode%
\setbox0=\hbox{$#1$}%
\setbox1=\hbox to\wd0{\hss$#2$\hss}\else%
\setbox0=\hbox{#1}%
\setbox1=\hbox to\wd0{\hss#2\hss}\fi%
 #1\hskip-\wd0\box1 }

\def\plusm#1#2{\buildrel {\raisebox{0.35ex}{\scriptsize $+#1$}} \over
{\raisebox{-0.35ex}{\scriptsize $-#2$}}}

\begin{document}

\hfill\vbox{\hbox{\bf NUHEP-TH-94-12}\hbox{May 1994}}\\

\vspace{0.5in}

\begin{center}
{\Large \bf Beauty Quark Fragmentation Into Strange B Mesons}\\

\vspace{0.4in}

Kingman Cheung\footnote{Internet address: {\tt cheung@nuhep.phys.nwu.edu}}
 and Robert J. Oakes\footnote{Internet address: {\tt oakes@fnalv.fnal.gov}}
 \\

\vspace{0.3in}

{\it Department of Physics and Astronomy, Northwestern University,\\
 Evanston, Illinois 60208, USA}
\end{center}

\vspace{0.5in}

\begin{abstract}

Using the recent measurement of the total production rate for $B_s$ and
$B_s^*$ mesons in electron-positron annihilation to determine the strange
quark mass parameter in the $\bar b\to B_s,\, B_s^*$ fragmentation functions
we calculate the momentum distributions of the $B_s$ and $B_s^*$ mesons.
\end{abstract}

\thispagestyle{empty}

\newpage

Recently the probability for a heavy quark to hadronize into a strange $B$ or
$D$ meson has been measured at LEP by the DELPHI Collaboration \cite{delphi}.
These data can be used to determine the strange quark mass parameter, $m_s$,
in the heavy quark fragmentation functions and, therefore, to predict the
momentum distributions of the $B_s$ and $B_s^*$ mesons produced in the
fragmentation process.  Using the heavy quark fragmentation functions
calculated in perturbative QCD at the scale of the heavy quark \cite{theory}
for the initial condition we  have numerically integrated the
Altarelli-Parisi equation to obtain the momentum distributions of the $B_s$
and $B_s^*$ mesons at the scale of the $Z$ mass.  The resulting distributions
are only moderately sensitive to the value of $m_s$, which we find to be
about 300 MeV$/c^2$.

The fragmentation function for the process $\bar b\to B_s$ at the scale of the
$b$ mass  is easily obtained  from the $\bar b\to B_c$ fragmentation functions
calculated using perturbative QCD by Braaten, Cheung, and Yuan
\cite{theory}:
\begin{eqnarray}
D_{\bar b\rightarrow B_s}(z,\mu_0) & = &
\frac{2\alpha_s(2m_s)^2 |R(0)|^2}
{81\pi m_s^3}\; \frac{rz(1-z)^2}{(1-(1-r)z)^6} \nonumber \\
&\times & [ 6 - 18(1-2r)z + (21 -74r+68r^2) z^2  \nonumber \\
 && -2(1-r)(6-19r+18r^2)z^3  + 3(1-r)^2(1-2r+2r^2)z^4 ]\,.
\label{dz1}
\end{eqnarray}
Here $z$ is the fraction of the $\bar b$ quark momentum carried by the $B_s$
meson,
$r=m_s/(m_b+m_s)$, and $R(0)$ is the $B_s$ meson $S$-wave radial  wavefunction
at the origin.  The corresponding fragmentation function for
$\bar b\to B_s^*$ is
\begin{eqnarray}
D_{\bar b\rightarrow B_s^*}(z,\mu_0) & = &
\frac{2\alpha_s(2m_s)^2 |R(0)|^2}
{27\pi m_s^3}\; \frac{rz(1-z)^2}{(1-(1-r)z)^6} \nonumber \\
&\times & [ 2 - 2(3-2r)z + 3(3 - 2r+ 4r^2) z^2 \nonumber \\
& &   -2(1-r)(4-r +2r^2)z^3  + (1-r)^2(3-2r+2r^2)z^4 ]\,,
\label{dz2}
\end{eqnarray}
The wavefunction at the origin, $R(0)$, can be determined from the $B_s$ meson
decay constant, $f_{B_s}$, from the Van Royen-Weisskopf \cite{weiss} relation,
modified for color:
\begin{equation}
\label{fbs}
f_{B_s}^2 = \frac{3}{\pi} \frac{|R(0)|^2}{M_{B_s}}\,.
\end{equation}
In Eqs.~(\ref{dz1}) and (\ref{dz2}) the running coupling constant
$\alpha_s(\mu)=\alpha_s(M_Z)/(1+b_0 \alpha_s(M_Z)\log(\mu/M_Z))$, where
$b_0=(33-2n_f)/6\pi$, $n_f$ is the number of flavors at the scale $\mu$,
$\alpha_s(M_Z)=0.12$, and we have chosen the scale to be $2m_s$ \cite{theory}.
The scale $\mu_0$ of the fragmentation functions is of the order of
$m_b$ and we choose it to be $m_b+2m_s$ \cite{theory}.
Integrating the Altarelli-Parisi equation for the
fragmentation functions from the initial scale $\mu_0$ to the $Z$ mass gives
the momentum distributions of the $B_s$ and $B_s^*$ mesons produced by the
fragmentation process in $e^+e^-$ annihilation at the $Z$ resonance.
To determine the strange
quark mass parameter, $m_s$, we use the recent measurement by the
DELPHI collaboration \cite{delphi} of the probability, $f^w_s$, that a weakly
decaying strange heavy meson is produced during the hadronization of a heavy
quark.  In addition to the direct production $\bar b\to B_s$, the
 $B_s$ meson can also result from the decays of excited states of $B_s$ and
$B_c$ mesons. However, these have been excluded in the measurement of $f_s^w$,
except for the radiative decay of the first excitation: $B_s^*\to B_s\; + \;
\gamma$.  Since the $B_s-B_s^*$ mass difference is so small we can
determine $m_s$ from the integrated fragmentation functions, which gives the
measured probability $f_s^w$:
\begin{equation}
\label{fws}
f^w_s = \int_0^1 dz \; \left[ D_{\bar b\to B_s}(z,\mu_0) +
D_{\bar b \to B_s^*}(z,\mu_0)
\right ]\,.
\end{equation}
Since the integrated fragmentation functions are independent of the scale, we
can choose the scale to be $\mu_0$ in Eq.~(\ref{fws}) and first determine
$m_s$ and then calculate the distributions $D_{\bar b\to B_s}(z,M_Z)$
and $D_{\bar b\to B_s^*}(z,M_Z)$ at the scale of the $Z$ mass
by numerically integrating  the Altarelli-Parisi equation.  As input we
fix the heavy $b$ quark mass at $m_b=5.0$ GeV$/c^2$.  To determine $R(0)$ we
use Eq.~(\ref{fbs}) with the value $f_{B_s}=207\pm 34\pm 22$ MeV from the
lattice calculations  of Bernard, Labrenz, and Soni \cite{lattice}.
Then from the measured value \cite{delphi}
\begin{equation}
f_s^w = 0.19 \pm 0.06 \pm 0.08
\end{equation}
we find (adding the two errors in quadrature)
\begin{equation}
m_s = 318 \plusm{47}{24} \;\; {\rm MeV}\,,
\end{equation}
which is in a reasonable range \cite{pdg}.    The
estimated error in $m_s$ reflects only the experimental uncertainty in
$f^w_s$.

Using this value of $m_s$ we have calculated the momentum distributions
$D_{\bar b\to B_s}(z)$ and $D_{\bar b\to B_s^*}(z)$,
which are shown in Figs.~\ref{fig1}.
Figure~\ref{fig1}(a) shows the fragmentation functions at the initial scale
$\mu_0=m_b+2m_s$, while Fig.~\ref{fig1}(b) shows the fragmentation functions
at the scale $\mu=M_Z$.    We have verified that the shapes of these
distributions are only moderately sensitive to the value of $m_s$,
although the peak does shift towards larger values of $z$ if $m_s$ is
decreased, as one might expect.  In fact, the maxima in both
$D_{\bar b\to B_s}(z,M_Z)$ and $D_{\bar b\to B_s^*}(z,M_Z)$ occur at
\begin{equation}
z_{\rm max}(B_s,\,B_s^*) = 0.92\,.
\end{equation}
The average momentum fraction $\langle z \rangle$ at the $Z$ mass scale is
also not too different for the $B_s$ and $B_s^*$ contributions to the spectrum:
\begin{equation}
\langle  z \rangle_{\bar b\to B_s}=0.65 \quad {\rm and} \quad
\langle  z \rangle_{\bar b\to B_s^*} =0.68\;.
\end{equation}
The spectrum also shows some evidence for approximate  heavy quark
spin symmetry at the level of about 50\%.  The ratio
$\int dz D_{\bar b\to B_s^*}/\int dz D_{\bar b\to B_s}$ is
about 2.1 compared to 3, as expected in the heavy quark limit from the spin
independence.  In terms of the ratio $P_V=B_s^*/(B_s+B_s^*)$, which heavy
quark symmetry predicts to be 0.75, we find 0.68.
To compare the $\bar b\to B_s$ and $\bar b\to B_s^*$ spectra individually
with experiments,  the two contributions $\bar b\to B_s$ and
$\bar b \to B_s^*\to B_s \gamma$ have to be experimentally separated by
observing the rather low energy photon, which has so far not been done.

To explore the sensitivity of our calculation to the assumptions we also
considered the fragmentation of a charm quark into a $D_s$ or $D_s^*$ meson.
Following ref.~\cite{delphi} we repeated the calculation assuming the
probability that a $D_s$ or $D_s^*$ meson is produced in the hadronization of
a $c$ quark is the same as the probability that a $B_s$ or $B_s^*$ meson is
produced in the hadronization of a $b$ quark; that is, $f^w_s$ is independent
of the heavy quark flavor.
Using the value  $f^w_s=0.19\pm0.06\pm0.08$ \cite{delphi}, the
lattice calculation $f_{D_s}=230\pm 30 \pm 18$~MeV \cite{lattice}, and choosing
$m_c=1.5$ GeV$/c^2$ \cite{pdg}, we found
\begin{equation}
\label{267}
m_s = 267 \plusm{36}{17} \;\; {\rm MeV}\,,
\end{equation}
which is in reasonable agreement with the value of $m_s$ obtained above
in the $B_s$ case.
With  this value of $m_s$ we also calculated the momentum spectra
for $c\to D_s$, for $c\to D_s^*$, and their sum at the initial scale
$\mu_0=m_c+2m_s$ as well as at the scale $\mu=M_Z$. They are shown,
respectively, in Figs.~\ref{fig2}(a) and \ref{fig2}(b).
As expected, the $D_s$  spectra are softer than the $B_s$ spectra shown in
Figs.~\ref{fig1}(a) and \ref{fig1}(b)  because
of the smaller charm quark mass.   The peaks of the spectra occur at 0.68
for $D_s$ and at 0.75 for $D_s^*$  and the mean momentum fractions are
$\langle z\rangle=0.49$ for
$D_s$ and $\langle z\rangle=0.52$ for $D_s^*$ at the scale $\mu=M_Z$
[Fig.~\ref{fig2}(b)].
The ratio of the two fragmentation probabilities for the $D_s$ case also
differs significantly more from the heavy quark spin symmetry prediction of 3,
 being about 1.6, than in the $B_s$ case, as one expects.  Also, the ratio
$P_V=D_s^*/(D_s+D_s^*)$ is 0.62, rather than 0.75.

To further test the sensitivity of our results to the input parameters assumed
we have repeated the calculation of the $D_s$ and $D_s^*$ spectra
using the measurement of the decay constant
$f_{D_s}$ recently reported by the CLEO collaboration \cite{cleo}.
Although this measured value $f_{D_s}=344 \pm 37 \pm 52 \pm 42$ MeV
\cite{cleo}  differs substantially from the lattice result
$f_{D_s}=230 \pm 30 \pm 18$  MeV \cite{lattice},
we found
\begin{equation}
m_s = 306 \plusm{45}{22} \;\; {\rm MeV}\,,
\end{equation}
which is not significantly different from the value above [Eq.~(\ref{267})],
and the changes in the fragmentation spectra shown in Fig.~\ref{fig2} are
negligible.

To summarize, we have calculated the momentum distributions of $B_s$ and
$B_s^*$ mesons produced by $\bar b$ quark fragmentation.  The measured total
production probability was used to determine the strange quark mass parameter.
The result for $m_s$ was quite reasonable, being well above the current quark
mass value, but lower than values often used for the constituent quark mass.
We note that at the scale $2m_s$ that $\alpha_s(2m_s)\approx 0.8$ is
uncomfortably large, which could mean there are significant corrections, which
have not been calculated, to the fragmentation function at the heavy quark
scale, which was calculated using perturbative QCD.
But this only affects the normalization and we determined the parameter $m_s$
{}from the experimental value of the total production probability.  Indeed, we
repeated the calculation for $D_s$ and $D_s^*$ meson
 production by fragmentation and
found substantially the same value for $m_s$ and verified that the momentum
distributions were only moderately sensitive to the value of $m_s$.   The
results were also not sensitive to the values assumed for the heavy quark
masses or the meson decay constants.
Essentially, the experimental production probability fixes the normalization,
the perturbative QCD result determines the shape of the momentum spectrum at
the heavy quark mass scale, and the Altarelli-Parisi equation governs the
evolution up to the scale $M_Z$; so perhaps it is not so surprising that the
results are not too sensitive to the particular values of the parameters.

\bigskip

This work was supported by the U.~S. Department of Energy, Division of
High Energy Physics, under Grant DE-FG02-91-ER40684.

\newpage
\begin{center}
\section*{Figure Captions}
\end{center}

\begin{enumerate}

\item
\label{fig1}
(a) Momentum distributions for the fragmentation processes
$\bar b \to B_s$,  $\bar b \to B_s^*$, and their sum at the scale
$\mu_0=m_b+2m_s$. \\
(b) Momentum distributions for the fragmentation processes
$\bar b \to B_s$,  $\bar b \to B_s^*$, and their sum at the scale
$\mu=M_Z$.

\item
\label{fig2}
(a) Momentum distributions for the fragmentation processes
$c \to D_s$,  $c \to D_s^*$, and their sum at the scale
$\mu_0=m_c+2m_s$. \\
(b)  Momentum distributions for the fragmentation processes
$c \to D_s$,  $c \to D_s^*$, and their sum at the scale
$\mu=M_Z$. \\

\end{enumerate}

\end{document}